\documentclass[sigconf,nonacm]{acmart}
\AtBeginDocument{%
  \providecommand\BibTeX{{%
    Bib\TeX}}}

\setcopyright{none}
\renewcommand{\footnotetextcopyrightpermission}[1]{}
\acmISBN{978-1-4503-XXXX-X/2018/06}

\usepackage{xcolor}
\usepackage{listings}
\usepackage{multicol}

\definecolor{dkgreen}{rgb}{0,0.6,0}
\definecolor{dkblue}{rgb}{0,0,0.6}
\definecolor{dkred}{rgb}{0.6,0,0}
\definecolor{ltblue}{rgb}{0,0.4,0.4}
\definecolor{dkviolet}{rgb}{0.3,0,0.5}

\lstdefinelanguage{Rocq}{
    mathescape=true,
    escapeinside={(*}{*)},
    texcl=false,
    morekeywords=[1]{Section, Module, End, Require, Import, Export,
        Variable, Variables, Parameter, Parameters, Axiom, Hypothesis,
        Hypotheses, Notation, Local, Tactic, Reserved, Scope, Open, Close,
        Bind, Delimit, Definition, Let, Ltac, Fixpoint, CoFixpoint, Add,
        Morphism, Relation, Implicit, Arguments, Unset, Contextual,
        Strict, Prenex, Implicits, Inductive, CoInductive, Record,
        Structure, Canonical, Coercion, Context, Class, Global, Instance,
        Program, Infix, Theorem, Lemma, Corollary, Proposition, Fact,
        Remark, Example, Proof, Goal, Save, Qed, Defined, Hint, Resolve,
        Rewrite, View, Search, Show, Print, Printing, All, Eval, Check,
        Projections, inside, outside, Def, MetaCoq, Run},
    morekeywords=[2]{forall, exists, exists2, fun, fix, cofix, struct,
        match, with, end, as, in, return, let, if, is, then, else, for, of,
        nosimpl, when},
    morekeywords=[3]{Type, Prop, Set, True, False, option},
    morekeywords=[4]{pose, set, move, case, elim, apply, clear, hnf,
        intro, intros, generalize, rename, pattern, after, destruct,
        induction, using, refine, inversion, injection, rewrite, congr,
        unlock, compute, ring, field, fourier, replace, fold, unfold,
        change, cutrewrite, simpl, have, suff, wlog, suffices, without,
        loss, nat_norm, assert, cut, trivial, revert, bool_congr, nat_congr,
        symmetry, transitivity, auto, split, left, right, autorewrite},
    morekeywords=[5]{by, done, exact, reflexivity, tauto, romega, omega,
        assumption, solve, contradiction, discriminate},
    morekeywords=[6]{do, last, first, try, idtac, repeat},
    morecomment=[s]{(*}{*)}, 
    showstringspaces=false,
    morestring=[b]", 
    morestring=[d],
    tabsize=3,
    extendedchars=false,
    sensitive=true,
    breaklines=false,
    basicstyle=\linespread{0.8}\footnotesize,
    captionpos=b,
    columns=[l]flexible,
    identifierstyle={\ttfamily\color{black}},
    keywordstyle=[1]{\ttfamily\color{dkviolet}},
    keywordstyle=[2]{\ttfamily\color{dkgreen}},
    keywordstyle=[3]{\ttfamily\color{ltblue}},
    keywordstyle=[4]{\ttfamily\color{dkblue}},
    keywordstyle=[5]{\ttfamily\color{dkred}},
    stringstyle=\ttfamily,
    commentstyle={\ttfamily\color{dkgreen}},
    literate=
    {\\forall}{{\color{dkgreen}{$\forall\;$}}}1
    {\\exists}{{$\exists\;$}}1
    {<-}{{$\leftarrow\;$}}1
    {=>}{{$\Rightarrow\;$}}1
    {==}{{$\approx\;$}}1
    {==>}{{$\geq\;$}}1
    {->}{{$\rightarrow\;$}}1
    {<->}{{$\leftrightarrow\;$}}1
    {<==}{{$\leq\;$}}1
    {\#}{{$^\star$}}1
    {\\o}{{$\circ\;$}}1
    {\/\\}{{\slash\symbol{92}}}1
    {\\\/}{{\textbackslash/}}1
    {~}{{$\sim$}}1
    {\@\@}{{$@$}}1
    {\\mapsto}{{$\mapsto\;$}}1
    {\\hline}{{\rule{\linewidth}{0.5pt}}}1 
}[keywords,comments,strings]

\lstdefinelanguage{Llvm}{
	sensitive=true,
	alsoletter={\%\#},
  escapeinside={(*}{*)},
	comment=[l]{;},
	string=[b]{"},
	keywords=[1]{
add, addrspacecast, alloca, and, ashr, atomicrmw, bitcast, br, call, cmpxchg,
extractelement, extractvalue, fadd, fcmp, fdiv, fence, fmul, fpext, fptosi,
fptoui, fptrunc, frem, fsub, getelementptr, icmp, indirectbr, insertelement,
insertvalue, inttoptr, invoke, landingpad, load, lshr, mul, or, phi, ptrtoint,
resume, ret, sdiv, select, sext, shl, shufflevector, sitofp, srem, store, sub,
switch, to, trunc, udiv, uitofp, unreachable, urem, va_arg, xor, zext
	},
	keywords=[2]{
acq_rel, acquire, addrspace, alias, align, alignstack, alwaysinline, any,
anyregcc, appending, arcp, arm_aapcs_vfpcc, arm_aapcscc, arm_apcscc, asm,
atomic, attributes, available_externally, blockaddress, builtin, byval, c,
catch, cc, ccc, cleanup, cold, coldcc, comdat, common, constant, datalayout,
declare, default, define, dereferenceable, dllexport, dllimport, eq, exact,
exactmatch, extern_weak, external, externally_initialized, false, fast, fastcc,
filter, gc, ghccc, global, hidden, inalloca, inbounds, initialexec, inlinehint,
inreg, intel_ocl_bicc, inteldialect, internal, jumptable, largest, linkonce,
linkonce_odr, localdynamic, localexec, max, min, minsize, module, monotonic,
msp430_intrcc, musttail, naked, nand, ne, nest, ninf, nnan, noalias, nobuiltin,
nocapture, noduplicate, noduplicates, noimplicitfloat, noinline, nonlazybind,
nonnull, noredzone, noreturn, nounwind, nsw, nsz, null, nuw, oeq, oge, ogt, ole,
olt, one, opaque, optnone, optsize, ord, personality, prefix, preserve_allcc,
preserve_mostcc, private, prologue, protected, ptx_device, ptx_kernel, readnone,
readonly, release, returned, returns_twice, samesize, sanitize_address,
sanitize_memory, sanitize_thread, section, seq_cst, sge, sgt, sideeffect,
signext, singlethread, sle, slt, spir_func, spir_kernel, sret, ssp, sspreq,
sspstrong, tail, target, thread_local, triple, true, type, ueq, uge, ugt, ule,
ult, umax, umin, undef, une, unnamed_addr, uno, unordered, unwind, uselistorder,
uselistorder_bb, uwtable, volatile, weak, weak_odr, webkit_jscc, x,
x86_64_sysvcc, x86_64_win64cc, x86_fastcallcc, x86_stdcallcc, x86_thiscallcc,
x86_vectorcallcc, xchg, zeroext, zeroinitializer
	},
	keywords=[3]{
i1, i2, i3, i4, i5, i6, i7, i8, i9, i10, i11, i12, i13, i14, i15, i16, i17, i18,
i19, i20, i21, i22, i23, i24, i25, i26, i27, i28, i29, i30, i31, i32, i33, i34,
i35, i36, i37, i38, i39, i40, i41, i42, i43, i44, i45, i46, i47, i48, i49, i50,
i51, i52, i53, i54, i55, i56, i57, i58, i59, i60, i61, i62, i63, i64, i80, i512,
void, half, float, double, fp128, x86_fp80, ppc_fp128, x86_mmx, label, metadata
	},
  commentstyle=\color{dkgreen},
	stringstyle=\color{red},
	keywordstyle=\color{dkblue},
	keywordstyle=[2]\color{purple},
	keywordstyle=[3]\color{dkred},
	basicstyle=\footnotesize\ttfamily,
	frame=lines,
	breaklines=true,
	prebreak=\raisebox{0ex}[0ex][0ex]{\ensuremath{\hookleftarrow}},
	showstringspaces=false,
	upquote=true,
	tabsize=3,
}[keywords,comments,strings]

\usepackage{textcomp}

\usepackage[framemethod=TikZ]{mdframed}

\mdfsetup{
  roundcorner=8pt,
  splittopskip=0,%
  splitbottomskip=0,%
  frametitleaboveskip=0,
  frametitlebelowskip=0,
  skipabove=0,%
  skipbelow=0,%
  innerleftmargin=1mm,%
  innerrightmargin=1mm,
  rightmargin=-1mm,%
  innertopmargin=-0.5mm,%
  innerbottommargin=-0.5mm,%
}
\lstnewenvironment{lf-yellow}{\mdframed}{\endmdframed}
\lstnewenvironment{lf-grey}{\mdframed[nobreak=true,linecolor=black!25,backgroundcolor=black!5]}{\endmdframed}

\usepackage{cleveref}

\begin{document}

\title{Towards Verified Compilation of Floating-point Optimization in Scientific Computing Programs}

\author{Mohit Tekriwal}
\email{tekriwal1@llnl.gov}
\affiliation{%
  \institution{Lawrence Livermore National Laboratory}
  \city{Livermore}
  \state{CA}
  \country{USA}
}

\author{John Sarracino}
\email{sarracino2@llnl.gov}
\affiliation{%
  \institution{Lawrence Livermore National Laboratory}
  \city{Livermore}
  \state{CA}
  \country{USA}
}

%
%
%
%

\renewcommand{\shortauthors}{Mohit Tekriwal \and John Sarracino}

\begin{abstract}
 Scientific computing programs often undergo aggressive compiler optimization to achieve high performance
and efficient resource utilization. While performance is critical, we also need to ensure that these
optimizations are correct. In this paper, we focus on a specific class of optimizations, floating-point
optimizations, notably due to fast math, at the LLVM IR level. We present a preliminary work, 
which leverages the Verified LLVM framework in the Rocq theorem prover, to prove the correctness
of Fused-Multiply-Add (FMA) optimization for a basic block implementing the arithmetic expression
$a * b + c$ . We then propose ways to extend this preliminary results by adding more program features
and fast math floating-point optimizations.
\end{abstract}

%

\keywords{ITrees, Verified LLVM, Floating-point, Optimization correctness}
%

\maketitle

\section{Introduction}
%
%
%
\noindent
Compiler optimization plays a crucial role in scientific computing by transforming source code into 
more efficient machine code, leading to significant performance improvements and efficient resource
utilization. Some common optimizations in high performance scientific computing codes include loop
optimizations such as loop unrolling, fusion and tiling, instruction level parallelism, dead code elimination,
vectorization and floating-point optimizations using fast math. We focus on fast math optimization, which are
compiler optimizations that speed up floating-point calculations by relaxing strict adherence to IEEE 754 standards.
These optimizations can significantly improve performance but may introduce subtle differences in results due to 
ordered operations, handling of special values such as NaN and Inf, and unsafe math optimizations. Several 
previous studies have documented correctness problems around numerical reproducibility and compiler
optimizations, including fused multiply-add (FMA)~\cite{guo2020pliner,ahn2021keeping,bentley2019multi}, flushing subnormal numbers to zero, clamping NaN (not-a-number)
to zero, and others. It is therefore important to verify that these optimizations are ``correct" or document 
conditions under which divergent numerical behaviors are noticed to help the scientific computing practitioners 
avoid such issues in the future. The goal of this work is to formally verify, in a \emph{mechanized setting},
 the correctness of floating-point optimizations. The approach we follow is that of translation validation~\cite{crellvm},
 where for each individual compilation pass, the optimizer generates a correctness proof justifying that particular run,
 and the proof checker validates the emitted proof. This proof checker is itself verified once-and-for-all, which results
 in an end-to-end assurance guarantee, even when using an otherwise untrusted compiler (such as LLVM). The core
 of the proof checker is a \emph{relational logic} which compares the optimized and the unoptimized code and verifies
 whether the translation of the unoptimized code to the optimized code is correct for some notion of correctness and 
 inference rules defined in this relational logic.
 
 Our work is inspired by Crellvm~\cite{crellvm}, which develops an infrastructure for translation validation of optimization passes
 for LLVM programs. This tool builds upon a verified LLVM framework (Vellvm)~\cite{vellvm}, implemented in the Rocq theorem prover~\cite{coq},
 which mechanizes the LLVM IR, its type system, denotations and operational semantics for LLVM programs and 
 reasoning principles for optimization passes. Crellvm however does not support reasoning about floating-point optimizations,
 and a direct extension of Crellvm for such optimizations is challenging because this tool has not been maintained for a long time (8 years) and 
 is therefore based on a very old version of Vellvm. The new version of Vellvm is significantly different from its old counterpart in two aspects - 
 language design and support for new version of LLVM (last checked, Vellvm supported LLVM 19.1.0). In this work, we build upon 
 the new version of Vellvm, to do a translation validation style proof of correctness of floating-point optimizations. We present preliminary
 result demonstrating steps for defining and proving a equivalence between two programs in Vellvm, which includes
 defining a denotation for these programs and refinement relation for the values returned by this program. These programs
define a FMA and a non-FMA implementation for the arithmetic expression $a * b + c$. The optimization correctness we consider
here is therefore that of FMA optimization. 
 
\section{Preliminaries}

\subsection{ITrees}
\noindent
Interaction trees (ITrees)~\cite{itrees} is a data structure, defined in the Rocq theorem prover, for representing computations
that can interact with an external environment. This structure is defined co-inductively to enable representing infinite
sequences of interactions or divergent behaviors, and is parameterized over \lstinline{E : Type -> Type}, which is a
type for external interactions, and \lstinline{R : Type} for the result type of the computation. This structure is built using
three constructors: \lstinline{Ret r}, which corresponds to trivial computation that immediately halts and returns a result $r$;
\lstinline{Tau t}, for representing silent computation with no visible event; and \lstinline{Vis A e k} constructor, which represents
outputs (of type A) provided by the computation done by a \emph{visible} event e, in response to its environment. $k$ represents
continuation or next computation in the sequence. The power of ITrees stems from the representation of program behavior as events,
and semantics for program behavior defined using semantics for each event handler. This allows us to decompose a program into 
a list of events and reason about each behavior modularly, and later compose them to prove correctness of the whole program.
In this work, we built our model program using three events: \emph{global event}, for global reads and writes, \emph{local event} for reading from and writing to local variables, and \emph{intrinsics event} for reasoning about the LLVM intrinsics.

\subsection{Vellvm}
\noindent
Verified LLVM (Vellvm)~\cite{vellvm} is a compiler framework, defined in the Rocq theorem prover, which mechanizes the 
LLVM Intermediate representation, LLVM type system, representation of LLVM programs using the ITree data structure,
and uses the monadic interpretation of ITrees to define and reason about a compositional, modular, and executable 
semantics for ``real-world" LLVM programs. A key powerful feature of Vellvm, which we use in our work, is its mechanism 
for verifying optimization passes, called \emph{equivalence up to taus (eutt}. Loosely speaking, \emph{eutt} relates two
programs $t_1$ with $t_2$, denoted as $t_1 \approx_R t_2$, if these programs (ITrees) are weakly bisimilar, i.e., if they 
produce the same tree of visible events, ignoring any finite number of Taus, or silent computations, and all values returned
along corresponding branches are related by $R$. Thus, the key to relating two programs is a careful definition of this 
refinement relation $R$. We demonstrate how to do that for our model problem in Section~\ref{sec:float_refine} and Section~\ref{sec:local_refine}.
Vellvm also provides a set of \emph{relational reasoning principles} that hold for $\approx_R$, which we use for proving correctness of
floating point optimization.
 
\subsection{Flocq}
\noindent
Flocq~\cite{boldo2011flocq} is a formal library defined in the Rocq theorem prover, which mechanizes the theory of IEEE-754 floating-point
arithmetic. This theory mechanizes both the standard $\delta-$ model, i.e., $a~ o_f~ b = (a~ o_R~ b) (1 + \delta)$ and the $\delta - \epsilon$
model, i.e.,   $a~ o_f~ b = (a~ o_R ~b) (1 + \delta) + \epsilon$, for error propagation in floating-point theory. This library defines the semantics 
for standard floating-point operations, such as $+, - , * , /, \sqrt{} $ and a set of lemmas for reasoning about the correctness of these 
operations. The power of this library is that the theory is parameterized in terms of the base radix, mantissa and exponent, and hence can used
to reason about any floating-point precision. Also, the library captures all the exceptional behavior of floating-point arithmetic through its 
inductive definition of a floating-point number. In this work, we base our proofs of floating-point optimizations on this library.

\section{Approach}

\subsection{Problem statement}
\noindent
In this preliminary work, we prove the equivalence between 
a basic block implementing the FMA optimization for computing
the arithmetic expression $a * b + c$ and the non-FMA version, 
which is implemented using two floating-point operations - 
product $p = a * b$ and the sum $p + c$. The implementation of
both these blocks are implemented in C at the source level and 
is then compiled to LLVM, as illustrated in Fig~\ref{fig:llvm_block}.

\lstset{
  language=Llvm,
  basicstyle=\footnotesize\ttfamily,
  columns=fullflexible,
  keepspaces=true,
  breaklines=true,
  frame=none,
  tabsize=2,
}
\begin{figure*}[h]
\begin{minipage}{.45\textwidth}
\begin{lf-grey}[language=Llvm]
define noundef double @f1(
  double noundef 
local_unnamed_addr (*\#*)0 {
    @llvm.fmuladd.f64 (double 
  ret double 
}
\end{lf-grey}
\end{minipage}
\hfill
\begin{minipage}{.45\textwidth}
\begin{lf-grey}[language=Llvm]
define noundef double @f1(
  double noundef 
local_unnamed_addr (*\#*)0 {
  ret double 
}
\end{lf-grey}
\end{minipage}
\vspace{-1em}
\caption{\footnotesize FMA v/s non-FMA code for the arithmetic expression $a * b + c$ in LLVM. 
This are the inputs to the Vellvm framework.}
\label{fig:llvm_block}
\vspace{-0.5em}
\end{figure*}

\lstset{
  language=Rocq,
  basicstyle=\footnotesize\ttfamily,
  columns=fullflexible,
  keepspaces=true,
  breaklines=true,
  frame=none,
  tabsize=2,
}

\noindent The LLVM blocks are then compiled using 
the Verified LLVM front end to obtain the corresponding
implementation the Rocq theorem prover. We then 
extract out only the block definition from the raw 
Rocq definition of the programs, as illustrated in Fig~\ref{fig:vellvm_block}.
At the moment, we do this manually, but our hope is 
that we can automate this process in the immediate 
follow up work. A block in Vellvm is defined as a record, with fields
representing the block id (\lstinline{blk_id}), phi nodes (\lstinline{blk_phis}), 
list of instructions defined inside the block (represented as \lstinline{blk_code}),
and returned value along with its type (represented as \lstinline{blk_term}).
As illustrated in Figure~\ref{fig:vellvm_block}, the FMA operation on the 
variables $a$, $b$ and $c$ is defined using an external call to
the LLVM intrinsic \lstinline{llvm.fmuladd.f64}. We define the semantics 
of this intrinsic in Rocq in Section~\ref{sec:intrinsics}. The non-FMA block on the 
other hand has two binary operations (\lstinline{OP_FBinop}) implemented, one
for multiplication, the semantics of which is implemented using the Flocq definition
of binary multiplication for floats (\lstinline{Bmult}), and the other for addition, which is 
implemented using Flocq's semantics for binary addition (\lstinline{Bplus}). An advantage
of using Flocq's definition for binary operations off the shelf, is that we can directly use 
the correctness lemmas associated with each floating-point unit. 

\begin{figure*}[htbp]
\footnotesize
\centering
\begin{minipage}{.45\textwidth}
\begin{lf-grey}
Definition fma_blk := {|
  blk_id := (Anon (3)
  blk_phis := [];
  blk_code := 
  [((IId (Anon (4)
    (INSTR_Call 
    (DTYPE_Double, 
    (EXP_Ident 
    (ID_Global (Name "llvm.fmuladd.f64")))) 
     [((DTYPE_Double, EXP_Double a), []);
       ((DTYPE_Double, EXP_Double b), []);
        ((DTYPE_Double, EXP_Double c), [])
     ]
     [ANN_tail Tail]
    ))
  ];
  blk_term := 
  (TERM_Ret (DTYPE_Double, 
    (EXP_Ident (ID_Local (Anon (4)
  blk_comments := None
  
|}.
\end{lf-grey}
\end{minipage}
\hfill
\begin{minipage}{.45\textwidth}
\begin{lf-grey}
Variable a b c : ll_double.
Definition non_fma_blk := {|
  blk_id := (Anon (3)
  blk_phis := [];
  blk_code := 
  [((IId (Anon (4)
    (INSTR_Op 
        (OP_FBinop FMul [] DTYPE_Double 
          (EXP_Double a)  (EXP_Double b)
     ))); 
    ((IId (Anon (5)
       (INSTR_Op (OP_FBinop FAdd [] DTYPE_Double 
        (EXP_Ident (ID_Local (Anon (4)
         (EXP_Double c) 
       )))
  ];
  blk_term := 
  (TERM_Ret (DTYPE_Double, 
       (EXP_Ident (ID_Local (Anon (5)
  blk_comments := None
|}.
\end{lf-grey}
\end{minipage}
\vspace{-1em}
\caption{\footnotesize Vellvm definition for FMA and non-FMA blocks}
\label{fig:vellvm_block}
\vspace{-1em}
\end{figure*}

The next step in this workflow is to inject these blocks into
Vellvm's ITree framework, which allows us to define equivalence 
between these two blocks using the powerful ``eutt" mechanism, 
and use the reasoning principles of ``eutt" off the shelf. This requires
us to define a denotation for these blocks in ITrees. We use the 
\lstinline{denote_block} function in ITrees, which is parameterized
in terms of the block, block id and its location in the memory. We instantiate
the block with \lstinline{non_fma_blk} and \lstinline{fma_blk} and their 
corresponding ids, to obtain denotations for the non-FMA and FMA block, respectively.
Since, we do not reason about memory events for now, we instantiate the
memory address to \lstinline{None}. Thus, the ITree programs corresponding
to the non-FMA and FMA blocks are as follows.

\noindent
\begin{minipage}{\columnwidth}
\begin{lf-grey}
Definition denote_nonfma_blk := 
  denote_block non_fma_blk (Anon (3)
  
Definition denote_fma_blk := 
  denote_block fma_blk (Anon (3)
\end{lf-grey}
\end{minipage}

Now that we have the ITree programs defined,
we can state the equivalence theorem.
The equivalence theorem is of the form:

\noindent
\begin{minipage}{\columnwidth}
\begin{lf-grey}
eutt equiv_rel ITree_prog1 ITree_prog2
\end{lf-grey}
\end{minipage}

\noindent
where \lstinline{equiv_rel} defines an equivalence relation 
between the returned state and the intermediate states of the programs \lstinline{ITree_prog1}
and \lstinline{ITree_prog2}. Since a state is defined as a map from a global or local variable to the value stored in these 
variables, the equivalence proof boils down to proving map equivalence for the two programs.  Since both the programs do not
modify the global environment,  the equivalence on global states
is trivial. The equivalence between local states boil down to proving that the values associated with the same local ids in 
both programs are equivalent for some definition of equivalence on the values, which we discuss in Section~\ref{sec:float_refine}. However,
since one local write in the FMA block is mathematically equivalent to two local writes in the non-FMA block, we need 
to define an alignment between the local write events. At the moment, we define this alignment relation manually, as discussed
in Section~\ref{sec:local_refine}. We expect to automate the calculation of this alignment relation in the future work. Once we have established
the equivalence relation on local states and the values stored in the local variables as well as the returned values, we can 
use these relations for equational reasoning in the ITree framework to prove the correctness of the FMA optimization or the 
equivalence between the FMA and non-FMA programs.

%

\subsection{Extending intrinsics in Vellvm with fmuladd}\label{sec:intrinsics}
\noindent
We add an intrinsic definition for FMA to Vellvm, depicted in \Cref{fig:vellvm_fma_intrinsic}.
This definition consists of two parts: a \emph{declaration} consisting of the intrinsic's LLVM function prototype 
(argument types, return type, etc), 
\verb|fmul_add_64|;
and a semantic definition of the function's behavior in Rocq, 
\verb|llvm_fmuladd_f64|.
Our semantic definition for FMA validates the input arguments
to ensure the correct function arity and that they have the correct dynamic type.
If so, it leverages a Compcert/Flocq definition for FMA, \verb|Bfma|
as the semantics for the FMA computation.
An advantage of this approach is that 
down the road, we will use Flocq's existing correctness lemma for FMA to reason about the
refinement relation between the FMA and non-FMA program, as we discuss in Section~\ref{sec:float_refine}.

\begin{figure*}[htbp]
\footnotesize
\centering
\begin{minipage}{.45\textwidth}
\begin{lf-grey}
Definition fmul_add_64 : declaration typ :=
 {|
   dc_name := Name "llvm.fmuladd.f64";
   dc_type := TYPE_Function TYPE_Double 
    [TYPE_Double; TYPE_Double; TYPE_Double] false;
    dc_param_attrs := ([], [[]]);
    dc_attrs       := [];
    dc_annotations  := []
 |}.
\end{lf-grey}
\end{minipage}
\hfill
\begin{minipage}{.45\textwidth}
\begin{lf-grey}
Definition llvm_fmuladd_f64 : semantic_function :=
fun args =>
 match args with
 |[DVALUE_Double a;DVALUE_Double b;DVALUE_Double c]=>
   let fma_op := 
   b64_fma DynamicValues.FT_Rounding a b c in
    ret (DVALUE_Double fma_op)
  | _ => 
   failwith "llvm_fmuladd_f64 got incorrect inputs"
 end.
\end{lf-grey}
\end{minipage}
\vspace{-1em}
\caption{\footnotesize Vellvm definition for FMA intrinsic.}
\label{fig:vellvm_fma_intrinsic}
\vspace{-0.5em}
\end{figure*}

\subsection{Refinement relation}
\noindent Key to proving the correctness of an optimization is a \emph{refinement relation} between 
LLVM programs.
A transformed program \emph{refines} the original program if the behaviors
of the original program include all the behaviors of the transformed program, i.e., $prog_1 \supseteq prog_2$.
In this preliminary work, the original program is the non-FMA block and the transformed program is the FMA block
with the behaviors between local reads and write, and FMA optimization of the arithmetic expression $a * b + c$.
In this section, we will explore the FMA optimization behavior, particularly, define what it means for an FMA program 
to refine its non-FMA counterpart in terms of the returned and intermediate values associated with local variables. 
Our refinement relation is given in \Cref{fig:refine_rel} and consists of two components,
\verb|double_refine|, a refinement relation between two LLVM values,
and \verb|local_refine|, a refinement relation between two local variable environments.
We discuss each in turn.

\subsection{Refinement relation for doubles}\label{sec:float_refine}

Vellvm has two denotations for values - \emph{dvalues}, also known as defined values or concrete values, and \emph{uvalues}, 
also known as under-defined values, which capture the non-deterministic behavior of LLVM programs due to interaction of operations
with \emph{undef}. By default, the refinement machinery in Vellvm allows one to define refinement relations over \emph{uvalues}.
Since, \emph{uvalues} are superset of \emph{dvalues} in Vellvm, Vellvm defines an automatic injection from defined values to 
under-defined values. So, in principle, one should be able to use the same refinement machinery over \emph{uvalues} for 
\emph{dvalues}. In this preliminary work, we define refinement relation over only defined values - \emph{double} and \emph{poison}.
The refinement relation \lstinline{double_refine} states that if both the concrete values $d_1$ and $d_2$ are poison, then $d_2$ refines $d_1$ 
if the type associated with the poison value is double in both cases, or if both the concrete values $d_1$ and $d_2$ are double then,
$d_2$ refines $d_1$ if $d_2$ is within an $\epsilon$ neighborhood of $d_1$. Since, we deal with floating-point values,
we cannot guarantee exact equivalence, hence the $\epsilon$ bound. In our case, this $\epsilon$ bound
will be the rounding error between the value from the FMA computation $d_2$ and the non-FMA computation, $d_1$, i.e.,
FMA and non-FMA computations are equivalent or FMA optimization is correct if $d_1$ and $d_2$ are equivalent modulo
rounding error. The specification further ensures that the values $p_1$, $p_2$ and the operation $p_1 \ominus_{RNE} p_2$ are finite, where we 
fix the rounding mode to RNE (rounding to nearest with ties to even), which is the default rounding mode.

\begin{figure*}
\begin{minipage}{.45\textwidth}
\begin{lf-grey}
Definition double_refine (d1 d2: uvalue):=
 match d1, d2 with
 | UVALUE_Poison t1, UVALUE_Poison t2 =>  
    dtyp_eqb t1 t2
 | UVALUE_Double p1, UVALUE_Double p2 => 
    is_finite p1 = true (*$\land$*)  is_finite p2 = true (*$\land$*)
    is_finite (p1 (*$\ominus_{RNE}$ *)p2) = true ->
    |p1 (*$\ominus_{RNE}$ *) p2| (*$\leq ~\epsilon_{FMA}$*)
 | _, _ => False
 end. 
\end{lf-grey}
\end{minipage}
\hfill
\begin{minipage}{.45\textwidth}
\begin{lf-grey}
Definition opt_double_refine x y := 
  match x, y with 
  | Some x_v, Some y_v => double_refine x_v y_v
  | _, _ => False
  end.

Definition local_refine (fma non_fma : local_env) := 
  opt_double_refine 
      (find (Anon 4) fma) 
      (find (Anon 5) non_fma) (* $\land$ *)
    remove (Anon 4) (remove (Anon 5) fma) = 
    remove (Anon 4) (remove (Anon 5) non_fma).
\end{lf-grey}
\end{minipage}
\vspace{-1em}
\caption{\footnotesize Refinement relations for Vellvm doubles and local environments.}
\label{fig:refine_rel}
\vspace{-0.5em}
\end{figure*}

Here, $\epsilon_{FMA} = (|(a * b * c)| \delta + \epsilon + |a * b| * (2\delta + \delta^2) + \epsilon  * (1 + \delta) + |c| * \delta + \epsilon) (1 + \delta) + \epsilon$, is a tight error bound for FMA operation. 
We derive this bound by collecting an upper bound on the absolute round off error associated with each operation, using the standard $\delta-\epsilon$ model in Flocq, and propagate through the expression. To generalize
the $\epsilon$ bound beyond FMA, we will need to symbolically track the error at each program
point. The existing Vellvm framework does not support such error tracking, and this will be our immediate focus
for future direction.

\subsection{Refinement relation for environments}\label{sec:local_refine}

The local variable relation needs to relate the effect of the two blocks on the respective local variables of each block.
In particular, the `eutt` relation applies the local variable relation to identical memories at the beginning of each block
and relates their output after executing the two blocks.
In this case, the ``FMA'' block writes `fma(a, b, c)' to the local variable \%4, 
while the non-``FMA'' block writes $a * b$ to \%4 and $\%4 + c$ to the local variable \%5.
Variables other than \%4 and \%5 are unmodified.

We define a local variable relation \verb|local_refine| that captures these semantics in two parts. 
First, the relation uses the \verb|double_refine| relation to relate the contents of \%4 on the FMA block and \%5 on the non-FMA block.
Second, we remove the variables added by the blocks (\%4 and \%5) and relate the remainders by strict equality.
Under the hood, Vellvm represents local variable environments as \emph{association lists} that map variables to `uvalues'. 
As a consequence, we implement these two constraints as `Prop's on the underlying association lists
by using association list `lookup` and `remove`.

\subsection{Closing the proof}
With the refinement relations for the values and the local environments defined,
we define an equivalence between the FMA and non-FMA block in Rocq in \Cref{fig:fma_optim_spec}.
The equivalence or FMA optimization correctness theorem 
\mbox{\lstinline{fma_optim_correct}} states that if the two 
programs 
\lstinline{interp_cfg2 denote_fma_blk g l} (ITree program for FMA block) and
\lstinline{interp_cfg2 denote_nonfma_blk g l} (ITree program for non-FMA block) start with the same
global environment $g$ and local environment $l$, the returned state of the the programs 
$(g_1, l_1, u_1)$ and $(g_2, l_2, u_2)$ are related by the refinement relation \\
\lstinline{(fun '(l1, (g1, u1)) '(l2, (g2, u2)) => ....}.  After desugaring the interpretation levels
for the ITree programs, where each level interprets a distinct interaction of the program with the 
environment, we apply the ITree relational reasoning principle  
\lstinline{eutt_Ret} to prove $(g_1, l_1, u_1) \approx    (g_2, l_2, u_2) $. The global environments
$g_1 = g_2 = g$, since both the programs do not modify the global environment. The local environments
are however modified, and we establish the equivalence between $l_1$ and $l_2$ using \lstinline{local_refine}, 
as discussed in Section~\ref{sec:local_refine}. Finally, we establish the equivalence between the concrete returned values
$u_1$ and $u_2$ using the refinement relation \lstinline{double_refine}, as discussed in Section~\ref{sec:float_refine},
 using the concrete round off error bound, $\epsilon_{FMA}$ between FMA and non-FMA, derived using the standard error model in Flocq and assuming the nice finiteness property on the input float values and the operations hold.

\begin{figure}
\begin{lf-grey}
Lemma fma_optim_correct: forall g l,
  eutt
  (fun '(l1, (g1, u1)) '(l2, (g2, u2)) =>
  	// Neither block changes globals  
  	g1 = g2 (* $\land$ *) 
  	local_refine l1 l2 (* $\land$ *)
  	// refinement relation for floats
  	match u1, u2 with
  	| inr u1_v, inr u2_v => 
      double_refine u1_v u2_v
  	| _, _ => False
  	end 
  )
  (interp_cfg2 denote_fma_blk g l) 
  (interp_cfg2 denote_nonfma_blk g l).
\end{lf-grey}
\vspace{-1em}
\caption{\footnotesize Equivalence between FMA and non-FMA blocks. }
\label{fig:fma_optim_spec}
\vspace{-1em}
\end{figure}

\section{Key takeaways and Future Direction}
\noindent
In this preliminary work, we demonstrated how one would go about proving correctness of an optimization pass
in Vellvm, by compiling an LLVM program in the Rocq theorem prover, defining denotation for program in a data structure
well suited for reasoning in Vellvm, defining refinement relations or meaning of optimization correctness, and finally proving
correctness of the program optimization with respect to the defined refinement relations, using the equational theory in Vellvm.
Owing to the foundational nature of Vellvm, we were able to use independent Rocq libraries like Flocq for floating-point reasoning.
An advantage of using this approach over other foundational tools like VST~\cite{vst}, is that this approach is source code agnostic.
As long as the source code compiles to an LLVM program, we should be able to verify correctness of the original code. VST on the
other hand is well suited for programs written in CompCert C. This makes using VST challenging for verifying correctness of scientific
programs, which are often written in Fortan, C++ and in some cases Julia and Python. Since all of these source languages are 
LLVM compatible, the Vellvm route offers the flexibility of reusing the correctness proofs on LLVM for any of these source languages.
That was our key motivation for using Vellvm for verifying the correctness of floating-point optimizations.

In the immediate future, we would like to extend this preliminary work in the following ways: \textbf{(a) parameterize the proofs over floating-point type}.
The FMA optimization proof we did here was only over double precision. In principle, we would like to prove the optimization correctness over
any floating-point type. \textbf{(b) Mechanize the semantics of other fast math optimizations}. We are working on extending the optimization correctness
proofs to other fast math optimizations such as Nnan, Ninf, Nsz, acrp, etc. \textbf{(c) Automate the proof}. The alignment predicate that we used
in this work to relate one computation of FMA with two computations of non-FMA was manually defined, owing to the simplicity of these computations.
However, as the complexity of computations increase, manually defining such alignment predicates becomes challenging. Therefore,
our immediate future work will be to instrument the compiler to compute these alignment predicates, which can then be used to set up refinement 
relations robustly. \textbf{(d) Add more program features}. Our preliminary work was over a program defining a basic block of computation. We are working
towards adding more program features such as loops, conditions, etc. and do the reasoning over whole control flow in the program.

This preliminary work was very promising in that it gives us a better idea of how to navigate a complex infrastructure such as Vellvm and identify
what components one would need to define to scale the verification to actual scientific programs.

%
\section*{Acknowledgements}
\noindent
This manuscript has been authored by Lawrence Livermore National Security, LLC under Contract No. DE-AC52-07NA27344 with
the U.S. Department of Energy. 
This material is
based upon work supported by the U.S. Department of Energy, Office of Science, Office of Advanced Scientific Computing Research
under the Correctness for Scientific Computing Systems (CS2) Program. LLNL IM number: LLNL-CONF-2009534.
\bibliographystyle{ACM-Reference-Format}
\bibliography{references}

\end{document}